\documentclass[intlimits,twoside,a4paper]{article}

\usepackage[cp1251]{inputenc}

\usepackage[eqsecnum]{cmpj3}

\usepackage{bm}
\issue{2020}{23}{1}{13601}
\doinumber{10.5488/CMP.23.13601}

\title[On the generalization of the Van der Waals approach]{On generalization of Van der Waals approach for isotropic-nematic fluid phase equilibria of anisotropic fluids in disordered porous medium}

\author[M.F. Holovko, V.I. Shmotolokha]{M.F. Holovko, V.I. Shmotolokha }
\address{
 Institute for Condensed Matter Physics of the National Academy of Sciences of Ukraine, 1 Svientsitskii St., 79011 Lviv, Ukraine
}

\date{Received July 26, 2019}

\begin{document}

\maketitle

\begin{abstract}
{A generalized Van der Waals approach is developed for anisotropic fluids in disordered porous media. As the reference system a hard spherocylinder fluid in a disordered porous medium is considered and described in the framework of the scaled particle theory with the Carnahan-Starling and Parsons-Lee corrections. The attractive part of interaction is treated in the framework of the mean field approximation in which due to orientational dependence of the excluded volume of two spherocylinders, a coupling between attractive and repulsive contributions is found. We focus on spherocylinder fluids with sufficiently long particle lengths for which the nematic-nematic transition was established. It is shown that these two nematic phases have different densities and are characterized by different orientational ordering. Strong influence of the type of interparticle attraction on the phase behaviour of anisotropic fluids in disordered porous media is established. Three simple models for this purpose are considered, namely a model with the Lennard-Jones anisotropic attraction, a model with modified Lennard-Jones attraction and a model with anisotropic square-well potential. For all considered models, the phase diagram shifts to the region of lower densities and lower temperatures as the porosity decreases.}
\keywords  {hard spherocylinder fluids, disordered porous media, scaled particle theory, generalized Van der Waals equation, isotropic-nematic equilibria, nematic-nematic transition}
%
\end{abstract}

\section{Introduction}
Since the discovery of liquid crystals by Planer \cite{Planer1961}, over 150 years ago and their rediscovery by Reinitzer \cite{Rein1888} in the late 19th century, liquid crystals have attracted attention due to the uniqueness of their thermodynamic, structural, optical and other properties. At present, one can recognize the liquid crystal behaviour in an ever-increasing number of scenarios: apart from the common examples of solutions of soaps and surfactants \cite{Chan92}, lyotropic liquid-crystalline order in biomacromolecular systems is ubiquitous in nature, including the phase behavior exhibited by DNA \cite{Nakata07}, by stiff polymers such as polysaccharides~\cite{Sato96}, cellulose \cite{Miller03} and protein fibers \cite{Mezzenga2010} and by rod-like viruses such as the tobacco mosaic virus \cite{Bawden36,Par76} or the fd virus \cite{Purdy05,Dogic06}. The supramolecular $\alpha$-helices formed from self-assembly of polypeptides in solutions are also found to give rise to a  rich variety of mesogenic behaviour \cite{Hor90,Tracy,Ginzburg03,Wu}.

An essential requirement for stabilization of a  liquid crystal phase is that the molecules should be highly anisotropic in shape, because they are widely used in hard sherocylinders for the description of isotropic-nematic transitions \cite{Vroege92}. This phase transition was first explained by Onsager \cite{Onsager49} seventy years ago as a result of competition between the orientational entropy that favors the disordered phase and the entropy effect associated with the orientational dependent excluded volume of spherocylinder-like particle that favours the order. In this approach, the molecular shape characterised by repulsive interactions is considered as the key to lyotropic liquid crystals in which the appearance of anisotropic phases is controlled by the solute concentration (or the density of anisotropic molecules).

It should be noted that the Onsager theory is based on the low-density expansion of the free  energy functional truncated at the second virial coefficient level and it is accurate for sufficiently long spherocylinders when the length of a spherocylinder 
$L_{1} \to\infty$ and the diameter $D_{1} \to\ 0$ in such a way that the non-dimensional concentration $C_{1}=\piup\rho_{1}L_{1}^{2}D_{1}/4$ is fixed,  where $\rho_{1}={N_{1}}/{V}$, $N_{1}$ is the number of spherocylinders, $V$ is the volume of the system \cite{Vroege92}. The application of the scaled particle theory (SPT) previously developed for a  hard-sphere fluid \cite{Reiss,Reiss1960} provides an  efficient way to incorporate higher  order contributions neglected in the Onsager theory \cite{Cotter70,Cotter74,Cotter70a}.

Another mechanism of formation of liquid crystalline matter can be connected with anisotropic attraction usually treated in molecular mean-field approaches such as the Maier-Saupe theory~\cite{Maier58,Maier1958}. In this approach, the orientationally dependent attractive interactions are considered as the key to the orientational order in thermotropic liquid crystals controlled by the temperature. In many cases, anisotropic fluids exhibit simultaneously lyotropic and thermotropic behaviour, which can be presented in concentration-temperature phase diagrams \cite{Wu,Franco-Melgar}. Due to this, both repulsive and attractive interactions between particles should be taken into account. This leads to the Van der Waals picture of fluids \cite{Van der Waals} in which the hard molecular core is treated as the reference system that determines the fluid structure while the attractions are incorporated by the perturbation way \cite{Barker1976,YukhHol80,Chandler}.
The generalized Van der Waals theory for anisotropic fluids was formulated by Cotter \cite{Cotter77,CotterM77,Cotter1977} and by the Gelbart group \cite{GelbartBaron,GelbartBar}.
 By combining the Onsager theory with the Van der Waals approach in the group of Jackson \cite{Wu,Franco-Melgar} for the attractive hard spherocylinders, four possible pairs of coexisting fluid phases  were predicted, namely vapor-liquid, vapor-nematic, liquid-nematic and nematic-nematic phases.

In our previous papers \cite{HolShmotPat,HolShmot2015}, the Van der Waals approach was generalized for the description of isotropic-nematic phase equilibria of  anisotropic fluids in a disordered porous medium. For that case, the Madden-Glandt model \cite{Madden} was used whereby a porous medium is presented as a quenched configuration of randomly distributed obstacles, for example the hard spheres in the simplest case. In accordance with the Van der Waals picture in the considered approach a hard spherocylinder fluid in a disordered porous medium is considered as the reference system. For the description of this reference system, the scaled particle theory has been used for the last decade extending the description of a  hard sphere fluid in a disordered porous medium \cite{HolDong,Chen2010,HolShmot2010,PatHolShmot2011,HolPatDong2012,HolPatDong2013,ChenZhao,HolPatDong2017} and generalized for the study of the influence of porous media on the isotropic-nematic transition in a hard spherocylinder fluid \cite{HolShmot2015,HolShmotPats2014,HolShmot2018} in disordered porous media and in hard spherocylinder-hard sphere mixture in bulk \cite{HolHvozd} and in porous media \cite{HvozdPatHol}.

However, in our previous papers \cite{HolShmotPat,HolShmot2015} for the treatment of attractive interaction in the generalized Van der Waals theory for anisotropic fluids in disordered porous media, we neglect the coupling between anisotropic repulsion and attractive parts in the anisotropic phase. In this paper we revise the theory presented in \cite{HolShmotPat,HolShmot2015} and analyze the coupling between anisotropic and attractive parts in the treatment of attractive interaction in the generalized Van der Waals equation for anisotropic fluids in disordered porous media. In addition, we will use our previous results \cite{HolShmot2018} for a hard spherocylinder fluid in a disordered porous medium obtained in the framework of the scaled particle theory with the Carnahan-Starling \cite{Carnahan69} and the Parsons-Lee \cite{Parsons79,Lee87} corrections in order to accurately describe the reference system at higher densities. In this paper, we focus on the consideration of anisotropic fluids with spherocylinders with rather large elongations for which in the nematic region a nematic-nematic transition is established. These two nematic phases have different densities and are characterized by different orientational ordering. We show that the phase diagram is very sensitive to the type of attractive inter-particle  interaction of the considered model. We show that a decrease of porosity shifts the nematic-nematic transition to lower densities and to lower temperatures.

\vspace{-2mm}

\section{The generalized Van der Waals theory}

As usual in the Van der Waals theory \cite{Van der Waals,Barker1976,YukhHol80,Chandler}, the expression for the thermodynamic properties of a fluid has two different contributions. The first one is connected with hard core repulsion which plays the role of the reference system in the description. The second contribution comes from the attractive part of interaction which is usually treated in a perturbation way. For example, the free energy of the fluid can be presented as the sum
\begin{equation}
\frac{F}{NkT}=\frac{F_{0}}{NkT}+\frac{F^\text{attr}}{NkT}\,,
\label{hol_smot2.1}
\end{equation}
where $N$ is the number of particles, $k$ is the Boltzmann constant, $T$ is the temperature, $F_{0}$ is the free energy of a  fluid of hard-body particles and, therefore, represents a purely repulsive contribution, $F^\text{attr}$ is the attractive part of the free energy.

\subsection{Thermodynamics of a hard spherocylinder fluid in random porous media. \\ Application of the scaled particle theory}

In this paper, as the reference system we consider a hard spherocylinder fluid in  random porous media created by the randomly distributed hard spheres. To describe the thermodynamic properties of this reference system we apply the scaled particle theory (SPT) developed for this purpose in our previous papers \cite{HolShmot2015,HolShmotPat,HolShmot2018}. According to the SPT approach, we introduce into the spherocylinder fluid in a random porous medium an additional spherocylinder with the scaling diameter $D_\text{s}$ and the scaling length $L_\text{s}$ as
\begin{equation}
D_\text{s}=\lambda_\text{s}D_{1}\,, \quad L_\text{s}=\alpha_\text{s}L_{1}\,,
\label{hol_smot2.2}
\end{equation}
where $D_{1}$ and $L_{1}$ are the diameter and the length of the fluid spherocylinder, respectively.

A key point of the SPT approach is based on the derivation of the chemical potential of this additional scaled particle and on the combination of the exact consideration of an infinitely small particle with thermodynamic consideration of a scaled particle of a sufficiently large size. The excess of chemical potential for the small scaled particle in a spherocylinder fluid in the presence of porous media can be written in the form \cite{HolShmot2015,HolShmotPat,HolShmot2018}
\begin{align}
\beta\mu_\text{s}^\text{ex}(\alpha_\text{s},\lambda_\text{s})&=-\ln
p_{0}(\alpha_\text{s},\lambda_\text{s})-\ln\left\{1-\frac{\eta_{1}}{V_{1}p_{0}(\alpha_\text{s},\lambda_\text{s})}\bigg[\frac{\piup}{6}D_{1}^{3}(1+\lambda_\text{s})^{3}\right.\nonumber\\
&+\frac{\piup}{4}D_{1}^{2}L_{1}(1+\lambda_\text{s})^{2}(1+\alpha_\text{s})\nonumber\\ 
&\left.+\frac{\piup}{4}D_{1}L_{1}^{2}(1+\lambda_\text{s})\alpha_\text{s}\int
f(\Omega_{1})f({\Omega_{2}})\sin\gamma(\Omega_{1},\Omega_{2})d\Omega_{1}d\Omega_{2}\bigg]\right\},
\label{hol_smot2.3}
\end{align}
where $\beta={1}/{kT}$, $\eta_{1}=\rho_{1}V_{1}$ is the fluid packing
fraction, $\rho_{1}={N_{1}}/{V}$ is the fluid density, $V_{1}$ is the spherocylinder volume, $V$ is the volume of the fluid,
\begin{equation}
p_{0}(\alpha_\text{s},\lambda_\text{s})=\exp\left[ -\beta\mu_\text{s}^{\circ}{(\alpha_\text{s},\lambda_\text{s})}\right] 
\label{hol_smot2.4}
\end{equation}
is the probability to find a cavity created by a scaled particle in the empty matrix. It is defined by the excess chemical potential $\mu_\text{s}^{\circ}{(\alpha_\text{s},\lambda_\text{s})}$ of the scaled particle in the limit of an infinite dilution of a fluid, $\Omega=(\vartheta, \varphi)$ denotes the orientation of particles which is defined by the angles $\vartheta$ and $\varphi$, $\rd \Omega=\frac{1}{4\piup}\sin\vartheta \rd\vartheta \rd \varphi$ is the normalized angle element, $\gamma(\Omega_{1}, \Omega_{2})$ is the angle between orientational vectors of two molecules; $f(\Omega)$ is the singlet orientational distribution function normalized in such a way that
\begin{equation}
\int f(\Omega)\rd \Omega = 1.
\label{hol_smot2.5}
\end{equation}
We note that hereafter we use conventional notations \cite{HolShmot2015,Madden,HolDong,Chen2010,HolShmot2010,PatHolShmot2011,HolPatDong2012,HolPatDong2013,ChenZhao,HolPatDong2017,HolShmotPats2014,HolHvozd,HolShmot2018}, where ``1'' is used to denote the fluid component, the index ``0'' denotes matrix particles, while for the scaled particles the index ``s'' is used.
For a large scaled particle, the excess chemical potential is given by a  thermodynamic expression for the work needed in order to create a macroscopic cavity inside a fluid and it can be presented in the form
\begin{equation}
\beta\mu_\text{s}^\text{ex}=W(\alpha_\text{s}\lambda_\text{s})+\beta P V_{s}/p_{0}(\alpha_\text{s}\lambda_\text{s})
\label{hol_smot2.6}\,,
\end{equation}
where $P$ is the pressure of the fluid, $V_\text{s}$ is the volume of the scaled particle.

In accordance with the ansatz of the SPT approach \cite{HolShmotPat,HolShmot2015} $W(\alpha_\text{s}\lambda_\text{s})$ can be presented in the form of expansion
\begin{equation}
W(\alpha_\text{s}\lambda_\text{s})=\omega_{00}+\omega_{10}\lambda_\text{s}+\omega_{01}\alpha_\text{s}+\omega_{11}\alpha_\text{s}\lambda_\text{s}+\frac12\omega_{20}\lambda_\text{s}^{2}\,.
\label{hol_smot2.7}
\end{equation}
The coefficients of this expansion can be found from the continuity of the excess chemical potential given by the expressions (\ref{hol_smot2.3}) and (\ref{hol_smot2.7}) as well as from the corresponding derivatives $\partial\mu_\text{s}^\text{ex}/\partial\lambda_\text{s}$, $\partial\mu_\text{s}^\text{ex}/\partial\alpha_\text{s}$, ${\partial^{2}\mu_\text{s}^\text{ex}}/{(\partial\lambda_\text{s}\partial\alpha_\text{s})}$ and $\partial^{2}\mu_\text{s}^\text{ex}/\partial\lambda^{2}_{s}$ at $\alpha_\text{s}=\lambda_\text{s}=0$. After setting $\alpha_\text{s}=\lambda_\text{s}=1$ we found the relation between the pressure $P$ and the excess chemical potential of a fluid
\begin{eqnarray}
\beta(\mu_{1}^\text{ex}-\mu_{1}^{0})&=&-\ln(1-\eta_{1}/\phi_{0})+A\left( \tau(f)\right) \frac{\eta_{1}/\phi_{0}}{1-\eta_{1}/\phi_{0}}\nonumber\\
&+&B\left( \tau(f)\right) \frac{(\eta_{1}/\phi_{0})^{2}}{(1-\eta_{1}/\phi_{0})^{2}}+\frac{\beta P}{\rho_{1}}\frac{\eta_{1}}{\phi}\,,
\label{hol_smot2.8}
\end{eqnarray}
where
\begin{eqnarray}
A(\tau(f)) &=&6+\frac{6\left(\gamma_{1}-1\right)^2\tau(f)}{3\gamma_{1}-1}-
\frac{p'_{0\lambda}}{\phi_0}\left[4+\frac{3\left(\gamma_{1}-1\right)^2\tau(f)}{3\gamma_{1}-1}\right] \nonumber\\
&-&\frac{p'_{0\alpha}}{\phi_0}\left(1+\frac{6\gamma_{1}}{3\gamma_{1}-1}\right)-\frac{p''_{0\alpha\lambda}}{\phi_0}-\frac{1}{2}
\frac{p''_{0\lambda\lambda}}{\phi_0}+2\frac{p'_{0\alpha}p'_{0\lambda}}{\phi_0^{2}}+\left(\frac{p'_{0\lambda}}{\phi_0}\right)^2,
\label{hol_smot2.9}
\end{eqnarray}
\begin{eqnarray}
B(\tau(f)) =\left(\frac{6\gamma_{1}}{3\gamma_{1}-1}-\frac{p'_{0\lambda}}{\phi_0}\right)\left[\frac{3\left(2\gamma_{1}-1\right)}{3\gamma_{1}-1}+
\frac{3\left( \gamma_{1}-1\right)^2\tau(f)}{3\gamma_{1}-1}-\frac{p'_{0\alpha}}{\phi_0}
-\frac{1}{2}\frac{p'_{0\lambda}}{\phi_0}\right],
\label{hol_smot2.10}
\end{eqnarray}
\begin{equation}
\tau(f)=\frac{4}{\piup}\int f (\Omega_{1}) f (\Omega_{2})\sin\vartheta_{12}\rd \Omega_{1}\rd \Omega_{2}\,,
\label{hol_smot2.11}
\end{equation}
\begin{equation}
\gamma_{1}=1+L_{1}/D_{1}\,,
\label{hol_smot2.12}
\end{equation}
$p'_{0\lambda}=\frac{\partial p_{0}(\alpha_\text{s},\lambda_\text{s})}{\partial\lambda_\text{s}}$,
$p'_{0\alpha}=\frac{\partial p_{0}(\alpha_\text{s},\lambda_\text{s})}{\partial\alpha_\text{s}}$,
$p''_{0\alpha\lambda}=\frac{\partial^{2} p_{0}(\alpha_\text{s},\lambda_\text{s})}{\partial\alpha_\text{s}\partial\lambda_\text{s}}$,
$p''_{0\lambda\lambda}=\frac{\partial^{2} p_{0}(\alpha_\text{s},\lambda_\text{s})}{\partial\lambda^{2}_{s}}$,
are the  corresponding derivatives at $\alpha_\text{s}=\lambda_\text{s}=0$.

We note that the probability $p_{0}(\alpha_\text{s},\lambda_\text{s})$ is related to two different types of porosity introduced by us in \cite{PatHolShmot2011,HolPatDong2012,HolPatDong2013}. The first one corresponds to geometric porosity
\begin{equation}
\phi_{0}=p_{0}(\alpha_\text{s}=\lambda_\text{s}=0)
\label{hol_smot2.13}
\end{equation}
characterizing the free volume of the fluid.

The second type of porosity corresponds to the case $\alpha_\text{s}=\lambda_\text{s}=1$ and leads to thermodynamic porosity
\begin{equation}
\phi=p_{0}(\alpha_\text{s}=\lambda_\text{s}=1)=\exp(-\beta\mu_{1}^{\circ})
\label{hol_smot2.14}
\end{equation}
defined by the excess chemical potential of a fluid particle $\mu_{1}^{\circ}$ in the limit of an infinite dilution. It characterizes the adsorption of the fluid in an empty matrix.

Using the Gibbs-Duhem equation which relates the pressure $P$ of a fluid to its total chemical potential $\mu_{1}=\mu_{1}^\text{id}+\mu_{1}^\text{ex}$, one derives the fluid compressibility in the following form
\begin{align}
\beta\left(\frac{\partial P}{\partial\rho_{1}}\right)_{T}&=\frac{1}{\left(1-\eta_{1}/\phi\right)}
+[1+A(\tau(f))]\frac{\eta_{1}/\phi_{0}}
{\left({1-\eta_{1}/\phi}\right)
	\left(1-\eta_{1}/\phi_{0}\right)}\nonumber\\\label{hol_smot2.15}
&+[A(\tau(f))+2B(\tau(f))]\frac{\left(\eta_{1}/\phi_{0}\right)^{2}}{\left(1-\eta_{1}/\phi\right)\left(1-\eta_{1}/\phi_{0}\right)^{2}}\\
&+
2B(\tau(f))\frac{\left(\eta_{1}/\phi_{0}\right)^{3}}{\left(1-\eta_{1}/\phi\right)\left(1-\eta_{1}/\phi_{0}\right)^{3}}\nonumber\,,
\end{align}
where $\mu_{1}^\text{id}$ is the ideal part of chemical
 potential of the fluid particle.
 
After integration of the relation  (\ref{hol_smot2.15}) over $\rho_{1}$ one obtains the expressions for the chemical potential and for the pressure in the SPT2 approach \cite{HolShmot2015,HolShmotPat, HolPatDong2012,HolPatDong2013,ChenZhao,HolShmotPats2014}. The obtained expressions are correct at small densities but at higher densities in accordance with (\ref{hol_smot2.15}) they have two divergences which appear at $\eta_{1}=\phi$ and $\eta_{1}=\phi_{0}$, respectively. Since $\phi<\phi_{0}$, the divergence for $\eta_{1}=\phi$ occurs at lower densities compared to the second one. From geometrical point of view, such a divergence should appear at higher densities at $\eta_{1}=\phi^{*}$,
which should be between $\phi$ and $\phi_{0}$
\begin{equation}
\phi<\phi^{*}<\phi_{0}\,.
\label{hol_smot2.16}
\end{equation}

Different corrections and improvements of the SPT2 approach were proposed in \cite{HolShmotPat,PatHolShmot2011,HolPatDong2013,HolPatDong2017}. In this paper, we dwell upon the SPT2b1 approximation which appears after replacing $\phi$ by $\phi_{0}$ everywhere in (\ref{hol_smot2.15}) except the first term and after removing the divergence $\eta_{1}=\phi$ in the corresponding expression for the chemical potential by expanding the logarithmic term via the following modification
\begin{equation}
-\ln\left(1-\eta_{1}/\phi\right) \approx\ln \left(1-\eta_{1}/\phi_{0}\right)+\frac{\eta_{1}(\phi_{0}-\phi)}{\phi_{0}\phi(1-\eta_{1}/\phi_{0})}\,.
\label{hol_smot2.17}
\end{equation}
The SPT2b1 approximation is accurate at small, intermediate and higher fluid densities. The expressions for the chemical potential and for the pressure within the SPT2b1 approximation can be presented in the following forms \cite{HolShmotPat,HolPatDong2017}
\begin{align}
&\beta\left(\mu_{1}^\text{ex}-\mu_{1}^{0}\right)^\text{SPT2b1}=\sigma(f)-\ln(1-\eta_{1}/\phi_{0})+\big[1+A(\tau(f))\big]\frac{\eta_{1}/\phi_{0}}{1-\eta_{1}/\phi_{0}}
+\frac{\eta_{1}(\phi_{0}-\phi)}{\phi_{0}\phi(1-\eta_{1}/\phi_{0})}
\nonumber\\
&
+\frac12\big[A(\tau(f))+2B(\tau(f))\big]\frac{(\eta_{1}/\phi_{0})^{2}}{(1-\eta_{1}/\phi_{0})^{2}}+\frac{2}{3}B(\tau(f))\frac{(\eta_{1}/\phi_{0})^{3}}
{(1-\eta_{1}/\phi_{0})^{3}}\,,
\label{hol_smot2.18}
\end{align}
\begin{align}
&\left(\frac{\beta P}{\rho_{1}}\right)^\text{SPT2b1}=\frac{1}{1-\eta_{1}/\phi_{0}}\frac{\phi_{0}}{\phi}+\left(\frac{\phi_{0}}{\phi}-1\right)
\frac{\phi_{0}}{\eta_{1}}\ln\left(1-\frac{\eta_{1}}{\phi_{0}}\right)\nonumber\\
&+\frac{A(\tau(f))}{2}\frac{\eta_{1}/\phi_{0}}{(1-\eta_{1}/\phi_{0})^{2}}+\frac{2B(\tau(f))}{3}\frac{(\eta_{1}/\phi_{0})^{2}}{(1-\eta_{1}/\phi_{0})^{3}}\,,
\label{hol_smot2.19}
\end{align}
where
\begin{eqnarray}
\sigma(f)=\int f(\Omega)\ln f(\Omega)\rd \Omega
\label{hol_smot2.20}
\end{eqnarray}
is the entropic term.

From  the thermodynamic relationship
\begin{equation}
\frac{\beta F}{V}=\beta\mu_{1}\rho_{1}-\beta P
\label{hol_smot2.21}
\end{equation}
we can obtain an expression for the free energy. The free energy of a confined fluid is \cite{HolShmotPat} 
\begin{eqnarray}
\frac{\beta F}{N}^\text{SPT2b1} &=&  \sigma(f)+\ln\frac{\eta_{1}}{\phi}-1
-\ln\left(1-\frac{\eta_{1}}{\phi_{0}}\right)+\left(1-\frac{\phi_{0}}{\phi} \right)\bigg[1+\frac{\phi_{0}}{\eta_1}\ln(1-\eta_1/\phi_{0})\bigg] \nonumber\\
&+& \frac{A(\tau(f))}{2} \frac{\eta_1/\phi_0}{1-\eta_1/\phi_0}+ \frac{B(\tau(f))}{3}
\left(\frac{\eta_1/\phi_0}{1-\eta_1/\phi_0}\right)^2.
\label{hol_smot2.22}
\end{eqnarray}

However, we should note that the SPT approach is not accurate enough for higher fluid densities, where the Carnahan-Starling (CS) correction \cite{Carnahan69} should be included. As a result, the equation of state can be presented in the form \cite{HolShmot2018}
%
\begin{eqnarray}
\frac{\beta P^\text{SPT2b1-CS}}{\rho_{1}}=\frac{\beta P^\text{SPT2b1}}{\rho_{1}}+\frac{\beta \Delta P^\text{CS}}{\rho_{1}},
\label{hol_smot2.23}
\end{eqnarray}
where the first term is given by equation (\ref{hol_smot2.19}) and the second term is the CS correction
\begin{eqnarray}
\frac{\beta \Delta P^\text{CS}}{\rho_{1}}=-\frac{\left(\eta_{1}/\phi_0\right)^3}{\left(1-\eta_{1}/\phi_0\right)^3}.
\label{hol_smot2.24}
\end{eqnarray}

Likewise, the chemical potential can be presented in the form
\begin{align}
\beta\mu_{1}^{0}=(\beta\mu_{1})^\text{SPT2b1}+\beta(\Delta\mu_{1})\,,
\label{hol_smot2.25}
\end{align}
where the correction $(\Delta\mu_{1})^\text{CS}$ can be obtained from the Gibbs-Duhem equation
\begin{equation}
(\beta\Delta\mu_{1})^\text{CS}=
\beta\int_{0}^{\eta_{1}}\frac{\rd \eta_{1}}{\eta_{1}} \left(\frac{\partial \Delta P}{\partial \rho_{1}}\right).
\label{hol_smot2.26}
\end{equation}
As a result,
\begin{equation}
(\beta\Delta\mu_{1})^\text{CS}=
\ln\left(1-\frac{\eta_{1}}{\phi_0}\right)+\frac{\eta_{1}/\phi_0}{1-\eta_{1}/\phi_0}-\frac{1}{2}\frac{(\eta_{1}/\phi_0)^{2}}{(1-\eta_{1}/\phi_0)^{2}}-\frac{(\eta_{1}/\phi_0)^{3}}{(1-\eta_{1}/\phi_0)^{3}}.
\label{hol_smot2.27}
\end{equation}

The free energy can also be presented in  the form
\begin{align}
\frac{\beta F_{0}}{N_1}=\frac{\beta F}{N_1}^\text{SPT2b1}+\frac{\beta F}{N_1}^\text{CS},
\label{hol_smot2.28}
\end{align}
where the first term is given by equation  (\ref{hol_smot2.22}) and  the second term can be found from the relation  (\ref{hol_smot2.21})
\begin{equation}
\left( \frac{\beta\Delta F}{N_{1}}\right) ^\text{CS}=
\ln(1-\eta_{1}/\phi_0)+\frac{\eta_{1}/\phi_0}{1-\eta_{1}/\phi_0}-\frac{1}{2}\frac{\left(\eta_{1}/\phi_0\right)^2}{\left(1-\eta_{1}/\phi_0\right)^2}.
\label{hol_smot2.29}
\end{equation}

\subsection{The contribution of attractive interactions}

The contribution of attractive interactions to thermodynamic properties of a fluid can be taken into account in the framework of the perturbation theory such as the Barker-Henderson theory \cite{Barker1976} or in the framework of optimized cluster expansions \cite{YukhHol80,Chandler}. However, the first term of the perturbation related to the so-called high temperature approximation (HTA) is identical in both approaches, and in the considered case for the free energy can be expressed as \cite{Franco-Melgar}
\begin{align}
\frac{\beta F^\text{attr}}{V}&=\frac{1}{2}\rho^{2}\int u^\text{attr}(r_{12},\Omega_{1},\Omega_{2})f(\Omega_{1})f(\Omega_{2})g_{2}^{0}(r_{12},\Omega_{1},\Omega_{2})\rd \bar{r}_{12}\rd \Omega_{1}\rd \Omega_{2}\,,
\label{hol_smot2.30}
\end{align}
where $u^\text{attr}(r_{12},\Omega_{1},\Omega_{2})$ is the attractive part of the interparticle interaction, $g_{2}^{0}(r_{12},\Omega_{1},\Omega_{2})$ is the pair distribution function of the reference system. 

Similar to \cite{Franco-Melgar}, we can  introduce the orientation-dependent contact distance $\sigma(\Omega_{1},\Omega_{2},\Omega_{r})$, where $\Omega_{1}$ and $\Omega_{2}$ are orientations of two particles $1$ and $2$ and $\Omega_{r}$ is the orientation of the intermolecular vector $\bar{r}_{12}$ between the centers of mass of these two particles. In terms of $\sigma(\Omega_{1},\Omega_{2},\Omega_{r})$, the repulsive part of interaction $u^\text{rep}(r_{12},\Omega_{1},\Omega_{2})$ for hard core particles can be represented in the form
\begin{align}
u^\text{rep}(r_{12},\Omega_{1},\Omega_{2})
=\left\{\begin{array}{ll}
\infty\quad {\text {for}}\quad r_{12}<\sigma(\Omega_{1},\Omega_{2},\Omega_{r})\\
0\,\,\quad {\text {for}}\quad r_{12}> \sigma(\Omega_{1},\Omega_{2},\Omega_{r})
\end{array}\right.
\label{hol_smot2.31}
\end{align}
and the pair distribution function of the reference system  can be approximated as the pair distribution function of the hard sphere fluid in a random porous medium with the same molecular volume as that of hard spherocylinders
\begin{align}
g_{2}^{0}(r_{12},\Omega_{1},\Omega_{2})\approx g_{2}^{0}\left( r_{12}/\sigma(\Omega_{1},\Omega_{2},\Omega_{r})\right) 
\label{hol_smot2.32}
\end{align}

As a result, the expression (\ref{hol_smot2.30}) for the attractive contribution can be presented in the form 
\begin{align}
\frac{\beta F^\text{attr}}{V}=-12\rho_{1}\beta\eta_{1}a\,,
\label{hol_smot2.33}
\end{align}
where 
\begin{align}
a=-\frac{1}{\phi_{0}V_{1}}\int f (\Omega_{1})f (\Omega_{2})u^\text{attr}(r_{12},\Omega_{1},\Omega_{2})g_{2}^{0}\left[ \frac{r_{12}}{\sigma\left(\Omega_{1},\Omega_{2},\Omega_{r}\right)}\right] 
r_{12}^{2}\rd r_{12}\rd \Omega_{2}\rd \Omega_{1}\rd \Omega_{r}.
\label{hol_smot2.34}
\end{align}
The factor $1/\phi_{0}$ excludes the volume occupied by matrix particles, $V_{1}$ is the volume of a particle, $\eta_{1}=\rho_{1}V_{1}$. In terms of parameter $a$, the equation of state and the chemical potential of the fluid have the typical Van der Waals form 
\begin{align}
\frac{\beta P}{\rho_{1}}=\frac{\beta P_{0}}{\rho_{1}}-12\beta\eta_{1}a\,,
\label{hol_smot2.35}
\end{align}
\begin{align}
\beta\mu_{1}=\beta\mu_{1}^{\circ}-24\beta\eta_{1}a\,.
\label{hol_smot2.36}
\end{align}
In this paper we present the attractive part of the interaction potential in the form
\begin{align}
&u^\text{attr}\left(r_{12},\Omega_{1},\Omega_{2}\right) \nonumber\\
&=\left\{\begin{array}{ll}
\Big\{u^\text{attr}_\text{iso}\left(\frac{r_{12}}{\sigma}\right)+u^\text{attr}_\text{aniso}\left[\frac{r_{12}}{\sigma\left(\Omega_{1},\Omega_{2},\Omega_{r}\right)}\right]\Big\}
[\epsilon_{0}+\epsilon_{2}P_{2}\left(\cos\vartheta_{12}\right)]\,,&
r_{12}\geqslant  \sigma(\Omega_{1},\Omega_{2},\Omega_{r})\,,\\
0\,,& r_{12}< \sigma(\Omega_{1},\Omega_{2},\Omega_{r})\,,
\end{array}\right.
\label{hol_smot2.37}
\end{align}
where $u^\text{attr}_\text{iso}\left(\frac{r_{12}}{\sigma}\right)$ is the isotropic part of attraction, $u^\text{attr}_\text{aniso}\left[\frac{r_{12}}{\sigma\left(\Omega_{1},\Omega_{2},\Omega_{r}\right)}\right]$ is the anisotropic part of attraction, $P_{2}\left(\cos\vartheta_{12}\right)$ is the second Legendre polynomial, $\vartheta_{12}$ is the angle between the principal axes of two interacting particles, $\epsilon_{0}$ and $\epsilon_{2}$ characterised the strengths of the corresponding isotropic and anisotropic attractive interactions.

Following the traditional scheme \cite{Franco-Melgar} and using a dimensionless intermolecular distance $r^{*}={r}/{\sigma\left(\Omega_{1},\Omega_{2},\Omega_{r}\right)}$ one obtains
\begin{equation}
a=a_\text{iso}+a_\text{aniso}\,,
\label{hol_smot2.38}
\end{equation} 
where
\begin{align}
a_\text{iso}&=-\frac{4\piup\sigma^{3}}{\phi_{0}V_{1}}\int_{0}^{\infty}g^\text{hs}_{2}\left(r^{*}\right)u^\text{attr}_\text{iso}\left(r^{*}\right)r^{*2}\rd r^{*}\nonumber\\
&\times\left[\epsilon_{0}+\epsilon_{2}\int f\left(\Omega_{1}\right) f\left(\Omega_{2}\right)P_{2}\left(\cos\vartheta_{12}\right)\rd\Omega_{1}\rd \Omega_{2}\right],
\label{hol_smot2.39}
\end{align}
\begin{align}
a_\text{aniso}&=-\frac{1}{\phi_{0}V_{1}}\int \rd\Omega_{1}\rd\Omega_{2}f\left(\Omega_{1}\right)f\left(\Omega_{2}\right)\left[\epsilon_{0}+\epsilon_{2}P_{2}\left(\cos\vartheta_{12}\right)\right] \nonumber\\
&\times V^\text{exc}_{1}\left(\Omega_{1},\Omega_{2}\right)3\int_{0}^{\infty}g^\text{hs}_{2}\left(r^{*}\right)u^\text{attr}_\text{aniso}\left(r^{*}\right)r^{*2}\rd r^{*}, 
\label{hol_smot2.40}
\end{align}
\begin{align}
V^\text{exc}_{1}\left(\Omega_{1},\Omega_{2}\right)&=\frac{1}{3}\int \rd\Omega_{2}\left[\sigma\left(\Omega_{1},\Omega_{2},\Omega_{r}\right)\right]^{3}=\frac{4}{3}\piup D^{3}_{1}\nonumber\\
&+2\piup D^{2}_{1}L_{1}+2D_{1}L^{2}_{1}\sin\vartheta\left(\Omega_{1},\Omega_{2}\right)
\label{hol_smot2.41}
\end{align}
is the excluded volume formed by two spherocylinders with the orientations $\Omega_{1}$ and $\Omega_{2}$. As we can see $a_\text{aniso}$ is proportional to the excluded volume $V_{1}^\text{exc}\left(\Omega_{1},\Omega_{2}\right)$ which appears due to the repulsive interaction. The coefficient near $V_{1}^\text{exc}\left(\Omega_{1},\Omega_{2}\right)$ has the form of the integral of $u_\text{aniso}^\text{attr}\left(r^{*}\right)$. It means that $a_\text{aniso}$ characterizes the coupling between the repulsive and attractive contributions.

Now, to calculate the parameter $a$ we should define the pair distribution function of a hard sphere fluid in a porous medium $g^\text{hs}_{2}\left({r}/{\sigma}\right)$ and the attractive part of the interaction potential $u^\text{attr}\left(r,\Omega_{1},\Omega_{2}\right)$. As the first step for the description of $g^\text{hs}_{2}\left({r}/{\sigma}\right)$, the interpolation scheme proposed in \cite{HvozdKalyu2017} can be used. In this scheme, the contact value obtained from the SPT theory \cite{KalyuHolPatCum2014} is combined with the analytical result for the pair distribution function of the hard-sphere fluid obtained in the Percus-Yevick approximation for the bulk case \cite{Wez63}. We consider the interaction potential $u^\text{attr}\left(r,\Omega_{1},\Omega_{2}\right)$ more in detail in the next section. However, in the Van der Waals approach the pair distribution function $g^{0}_{2}\left(r_{12},\Omega_{1},\Omega_{2}\right)$ is approximated by its low-density limit \cite{Franco-Melgar}
\begin{equation}
\lim_{\rho_{1}\rightarrow 0}g^{0}_{2}\left(r_{12},\Omega_{1},\Omega_{2}\right)=\exp\left[-\beta u^\text{rep}\left(r_{12},\Omega_{1},\Omega_{2}\right)\right],
\label{hol_smot2.42}
\end{equation}
where the repulsive part of the potential $u^\text{rep}\left(r_{12},\Omega_{1},\Omega_{2}\right)$ is defined by equation (\ref{hol_smot2.31}).

As a result, in the Van der Waals approach the calculation of the constants $a_\text{iso}$ and $a_\text{aniso}$ reduces to the calculations of the mean values of the potentials $u^\text{attr}_\text{iso}\left(r^{*}\right)$ and $u^\text{attr}_\text{aniso}\left(r^{*}\right)$, correspondingly. We note that the background of the Van der Waals approach is connected with the consideration of the potentials $u^\text{attr}_\text{iso}\left(r^{*}\right)$ and $u^\text{attr}_\text{aniso}\left(r^{*}\right)$ in the form of Kac potentials $u^\text{attr}_\text{iso}\left(r^{*}\right)=\gamma^{3}\Phi^\text{attr}_\text{iso}\left(\gamma r\right)$ and $u^\text{attr}_\text{aniso}\left(r^{*}\right)=\gamma^{3}\Phi^\text{attr}_\text{aniso}\left(\gamma r\right)$ whose range ${1}/{\gamma}$ is very long compared to the range of the repulsive potential $u^\text{rep}\left[{r}/{\sigma\left(\Omega_{1},\Omega_{2},\Omega_{r}\right)}\right]$ and in the limit $\gamma\rightarrow 0$ the expressions (\ref{hol_smot2.33}),(\ref{hol_smot2.35}),(\ref{hol_smot2.36}) in the Van der Waals approximation (\ref{hol_smot2.42}) will be exact \cite{KacUhlenHam63,LebowStellBaer65}.

However, as the next step of the calculations we need to know the singlet distribution function $f\left(\Omega\right)$ which can be found from the minimization of the total free energy of the fluid presented as the sum (\ref{hol_smot2.1}). Such a minimization leads to the integral equation for the distribution function $f\left(\Omega\right)$.

\subsection{The integral equation for the singlet distribution function}

The minimization of the total free energy with respect to $f\left(\Omega\right)$ leads to a nonlinear integral equation for the singlet distribution function
\begin{align}
&\ln f\left(\Omega_{1}\right)+\lambda+\frac{8}{\piup}C\int f\left(\Omega_{2}\right)\sin\vartheta_{12}\rd\Omega_{2}+\beta\rho_{1}\epsilon_{0}\frac{4\piup\sigma^{3}}{\phi_{0}}\int_{1}^{\infty}u^\text{attr}_\text{iso}\left(r^{*}\right)r^{*2}\rd r^{*}\nonumber\\
&\times\int f\left(\Omega_{2}\right)P_{2}\left(\cos\vartheta_{12}\right)\rd\Omega_{2}+\beta\rho_{1}\epsilon_{2}\frac{8}{\piup}L^{2}_{1}D_{1}\frac{1}{\phi_{0}}\nonumber\\
&3\int_{1}^{\infty}u^\text{attr}_\text{aniso}\left(r^{*}\right)r^{*2}\rd r^{*}\int f\left(\Omega_{2}\right)
\sin\vartheta_{12}P_{2}\left(\cos\vartheta_{12}\right)d\Omega_{2}=0\,,
\label{hol_smot2.43}
\end{align}
where the constant $\lambda$ can be found from the normalization condition (\ref{hol_smot2.5}).

The constant $C$ can be presented as the sum
\begin{equation}
C=C_\text{rep}+C_\text{attr}\,,
\label{hol_smot2.44}
\end{equation}
where
\begin{equation}
C_\text{rep}=\frac{{\eta_{1}}/{\phi_{0}}}{1-{\eta_{1}}/{\phi_{0}}}\left[\frac{3\left(\gamma_{1}-1\right)^{2}}{3\gamma_{1}-1}\right]\left(1-\frac{P^{'}_{0\lambda}}{2\phi_{0}}\right)+\frac{{\eta_{1}}/{\phi_{0}}}{\left(1-{\eta_{1}}/{\phi_{0}}\right)}\delta\left(\frac{6\gamma_{1}}{3\gamma_{1}-1}-\frac{P^{'}_{0\lambda}}{\phi_{0}}\right)
\label{hol_smot2.45}
\end{equation}
is the contribution from the repulsive interaction part. The constant $\delta={3}/{8}$ is the Parsons-Lee \cite{Parsons79,Lee87} correction introduced by us in \cite{HolShmot2018}.
\begin{equation}
C_\text{attr}=\beta\rho_{1}\epsilon_{0}\frac{8}{\piup}L_{1}^{2}D_{1}\frac{1}{\phi_{0}}3\int_{1}^{\infty}u^\text{attr}_\text{aniso}\left(r^{*}\right)r^{*2}\rd r^{*}
\label{hol_smot2.46}
\end{equation}
is the contribution from the attractive part of interaction.

Instead of the solution of the integral equation (\ref{hol_smot2.43}), in literature the researchers usually  used for $f\left(\Omega\right)$ the trial function in the Onsager \cite{Onsager49}, Odijk \cite{Odijk86} or other forms with parameters calculated from the minimization of the free energy. Such a procedure leads to algebraic equations but it usually overestimates the orientational ordering in the fluid \cite{Vroege92}. Instead of such a procedure, in this paper we solve the integral equation (\ref{hol_smot2.43}) using the scheme presented in \cite{HerBerWin84}.

We note that in the case when $\epsilon_{2}=0$ the equation (\ref{hol_smot2.43}) has the same structure as the corresponding equation
\begin{equation}
\ln f\left(\Omega_{1}\right)+\lambda+\frac{8}{\piup}C\int f\left(\Omega_{2}\right)\sin\vartheta_{12}\rd\Omega_{2}=0
\label{hol_smot2.47}
\end{equation}
obtained by Onsager \cite{Onsager49} for a hard spherocylinder fluid in the limit $L_{1}\rightarrow \infty,D_{1}\rightarrow 0$ while the dimensionless density of the fluid $c=\frac{1}{4}\piup\rho_{1}L_{1}^{2}D_{1}$ is fixed. In this limit $C\rightarrow c$. From the bifurcation analysis of the equation (\ref{hol_smot2.47}), the existence of two characteristic points was found \cite{KayRav78}
\begin{equation}
C_\text{i}=3.290, \quad C_\text{n}=4.191
\label{hol_smot2.48}
\end{equation}
which define the high density of a stable isotropic fluid and the minimal density of a stable fluid in the nematic state.

In accordance with (\ref{hol_smot2.44}), the constant $C$ is defined by the sum of two terms. The first term due to (\ref{hol_smot2.45}) depends on the packing fraction $\eta_{1}$ and has a positive value. The second term $C_{2}$ describes the attractive contribution and is proportional to the inverse temperature $\beta={1}/{kT}$. This term has a negative value.

\section{Results and discussions}

In this section we apply the theory developed in the previous section for the description of the phase behaviour of anisotropic fluids in disordered porous media at a fixed value $\gamma_{1}=1+{L_{1}}/{D_{1}}$. The calculated phase diagrams are presented in terms of dimensionless variables: $T^{*}=({kT})/{\epsilon_{0}}$ for the temperature and $\eta_{1}=\rho_{1}V_{1}$ for the packing fraction. The richness of nematic-liquid-vapour phase behaviour for a hard spherocylinder fluid with an attractive interparticle interaction was demonstrated for the bulk case \cite{Franco-Melgar} and in porous media \cite{HolShmotPat,HolShmot2015}. One of the most striking features of this phase behaviour is the appearance of a region of nematic-nematic phase separation for high particle elongations, namely at ${L_{1}}/{D_{1}}>50$. We note that in the Onsager limit when $L_{1}\rightarrow\infty, D_{1}\rightarrow 0$ and $C_{1}=\frac{1}{4}\piup\rho_{1}L_{1}^{2}D_{1}$ is fixed, $\eta_{1}=\rho_{1}V_{1}\rightarrow 0$ since in this limit $V_{1}\rightarrow 0$. The nematic-nematic region involves the coexistence of a low-density vapor-like anisotropic state of lower orientational order with a high-density liquid anisotropic state of higher orientational order. As for the usual vapor-liquid transition, both phases are bounded by the critical point at higher temperatures which exists in the nematic region.

In this paper we focus on the influence of the types of interparticle attraction and the presence of porous media on the nematic-nematic phase separation. As mentioned recently in \cite{Wu}, the position of the isotropic-nematic transition is determined principally by the length of spherocylinders, and coexistence between the isotropic and the low-density nematic phases $N_{1}$ is not affected by incorporation of anisotropy into the attractive interactions. The enhanced anisotropic attractive interaction only shifts the nematic-nematic coexistence curves to higher temperatures and also shifts the isotropic-nematic-nematic triple point temperature to a higher temperature. However, the form of the phase diagram does not change. Due to this, similar to \cite{Wu}, here for simplification we put $\epsilon_{2}=0$. In this case, we come back to the integral equation (\ref{hol_smot2.43}) for the singlet distribution function, and the coexistence curves can be found from the conditions of thermodynamic equilibrium
\begin{align}
	&\mu_{1}\left(\rho_{1}^{1},T\right)=\mu_{1}\left(\rho_{1}^{2},T\right),\nonumber\\
	&P\left(\rho_{1}^{1},T\right)=P\left(\rho_{1}^{2},T\right),
\label{hol_smot3.1}
\end{align}  
where $\mu_{1}\left(\rho_{1}^{1},T\right)$ and $P\left(\rho_{1}^{1}T\right)$ are the chemical potential and the pressure of the fluid correspondingly, $\rho_{1}^{1}$ and $\rho_{1}^{2}$ are the fluid densities of two different phases $1$ and $2$. The numerical solution of the equations (\ref{hol_smot3.1}) is realized using the Newton-Raphson algorithm.

\begin{figure}[!b]
	\centerline{
				\includegraphics [height=0.45\textwidth]{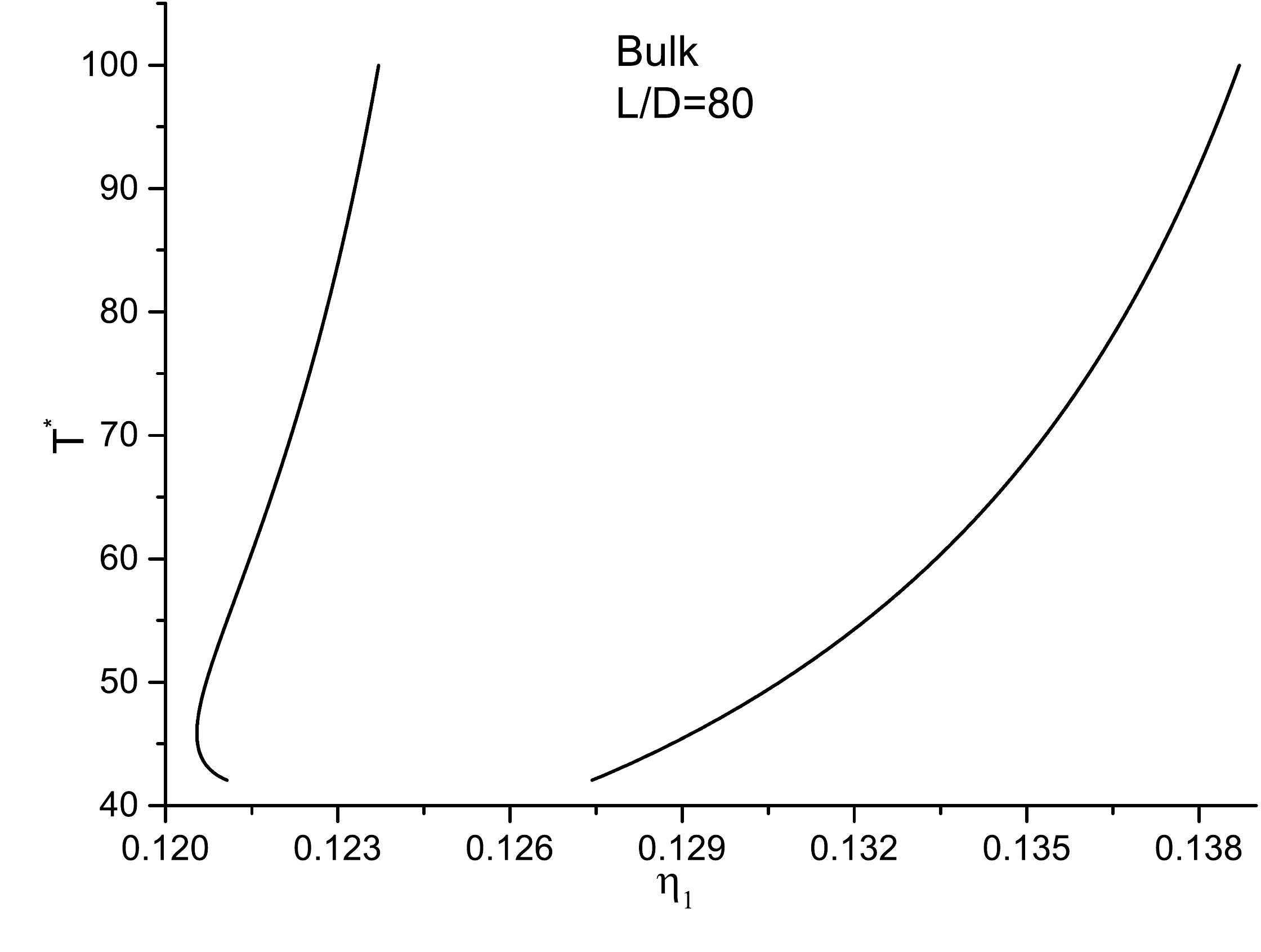}
	}
	\caption{Phase diagram for a hard-spherocylinder fluid with an anisotropic Lennard-Jones attraction.}
	\label{Fig1}
\end{figure}

We start our investigation from the simple model of a hard spherocylinder fluid with an attractive interaction in the form (\ref{hol_smot2.37}), in which $u_\text{iso}^\text{attr}\left(\frac{r}{\sigma}\right)=0$ and $u_\text{aniso}^\text{attr}\left[\frac{r}{\sigma\left(\Omega_{1},\Omega_{2},\Omega_{r}\right)}\right]$ has the Lennard-Jones-like form

\begin{align}
&u_\text{aniso}^\text{attr}\left[\frac{r}{\sigma\left(\Omega_{1},\Omega_{2},\Omega_{r}\right)}\right]\nonumber\\
&=\left\{\begin{array}{ll}
4\epsilon_{0}\left[\left(\frac{\sigma(\Omega_{1},\Omega_{2},\Omega_{r})}{r}\right)^{12}-
\left(\frac{\sigma(\Omega_{1},\Omega_{2},\Omega_{r})}{r}\right)^{6}\right],&
r>  \sigma(\Omega_{1},\Omega_{2},\Omega_{r})\,,\\
0\,,& r< \sigma(\Omega_{1},\Omega_{2},\Omega_{r})\,.
\end{array}\right.
\label{hol_smot3.2}
\end{align}

For simplification we put $\epsilon_{2}=0$. For the case considered
\begin{equation}
	\int_{1}^{\infty}u_\text{aniso}^\text{attr}\left(r^{*}\right)r^{*2}\rd r^{*}=-\frac{8}{9}\,.
\label{hol_smot3.3}
\end{equation} 
The phase diagram for a hard spherocylinder fluid with the Lennard-Jones-like attraction is presented in figure \ref{Fig1} for the bulk case at ${L_{1}}/{D_{1}}=80$.
We remember that the potential (\ref{hol_smot3.2}) was used by us in our previous papers \cite{HolShmotPat,HolShmot2015}, in which, however, for simplification the excluded volume $V_{1}^\text{exc}\left(\Omega_{1},\Omega_{2}\right)$ in the expression (\ref{hol_smot2.40}) was approximated by its value for the isotropic case
\begin{equation}
	V_\text{1,iso}^\text{exc}\left(\Omega_{1},\Omega_{2}\right)=\frac{4}{3}\piup D_{1}^{3}+2\piup D_{1}^{2}L_{1}+\piup D_{1}L_{1}^{2}\,.
\label{hol_smot3.4}
\end{equation}
The comparison of figure \ref{Fig1} with the corresponding results obtained in \cite{HolShmotPat,HolShmot2015} in the framework of approximation (\ref{hol_smot3.4}) demonstrates that to calculate the attractive interaction part, the inclusion of the non-approximated expression (\ref{hol_smot2.41}) for the excluded volume formed by two spherocylinders is very important and can  completely change the phase diagram. Probably Lennard-Jones-like model overestimates the role of anisotropy for interparticle attraction at large distances. Due to this, we introduce the second model which is a slightly modified version of the original Lennard-Jones model. In this model we cut attraction between particles at $r^{*}=\gamma_{1}=1+{L_{1}}/{D_{1}}$ and we shift the interaction potential to the value $u_\text{aniso}^\text{attr}\left(r^{*}=\gamma_{1}\right)$. Consequently, in the second model
\begin{align}
u_\text{aniso}^\text{attr}\left[\frac{r}{\sigma\left(\Omega_{1},\Omega_{2},\Omega_{r}\right)}\right]=\left\{\begin{array}{ll}
4\left[\frac{1}{\left(r^{*}\right)^{12}}-\frac{1}{\left(r^{*}\right)^{6}}\right]-4\left(\frac{1}{\gamma_{1}^{12}}-\frac{1}{\gamma_{1}^{6}}\right), & \quad
1<r^{*}<\gamma_{1}\,,\\
0, &\quad r^{*}<1, \,\, r^{*}>\gamma_{1}\,.
\end{array}\right.
\label{hol_smot3.5}
\end{align}

As a result, instead of (\ref{hol_smot3.3}) we have
\begin{equation}
\int_{1}^{\infty}u_\text{aniso}^\text{attr}\left(r^{*}\right)r^{*2}\rd r^{*}=-\frac{8}{9}+\frac{4}{3\gamma_{1}^{3}}\left(1-\frac{1}{3\gamma_{1}^{6}}\right)+\frac{4}{3\gamma_{1}^{6}}\left(\gamma_{1}^{3}-1\right)\left(1-\frac{1}{\gamma_{1}^{6}}\right).
\label{hol_smot3.6}
\end{equation}
\begin{figure}[!b]
	\centerline{
				\includegraphics [height=0.45\textwidth]{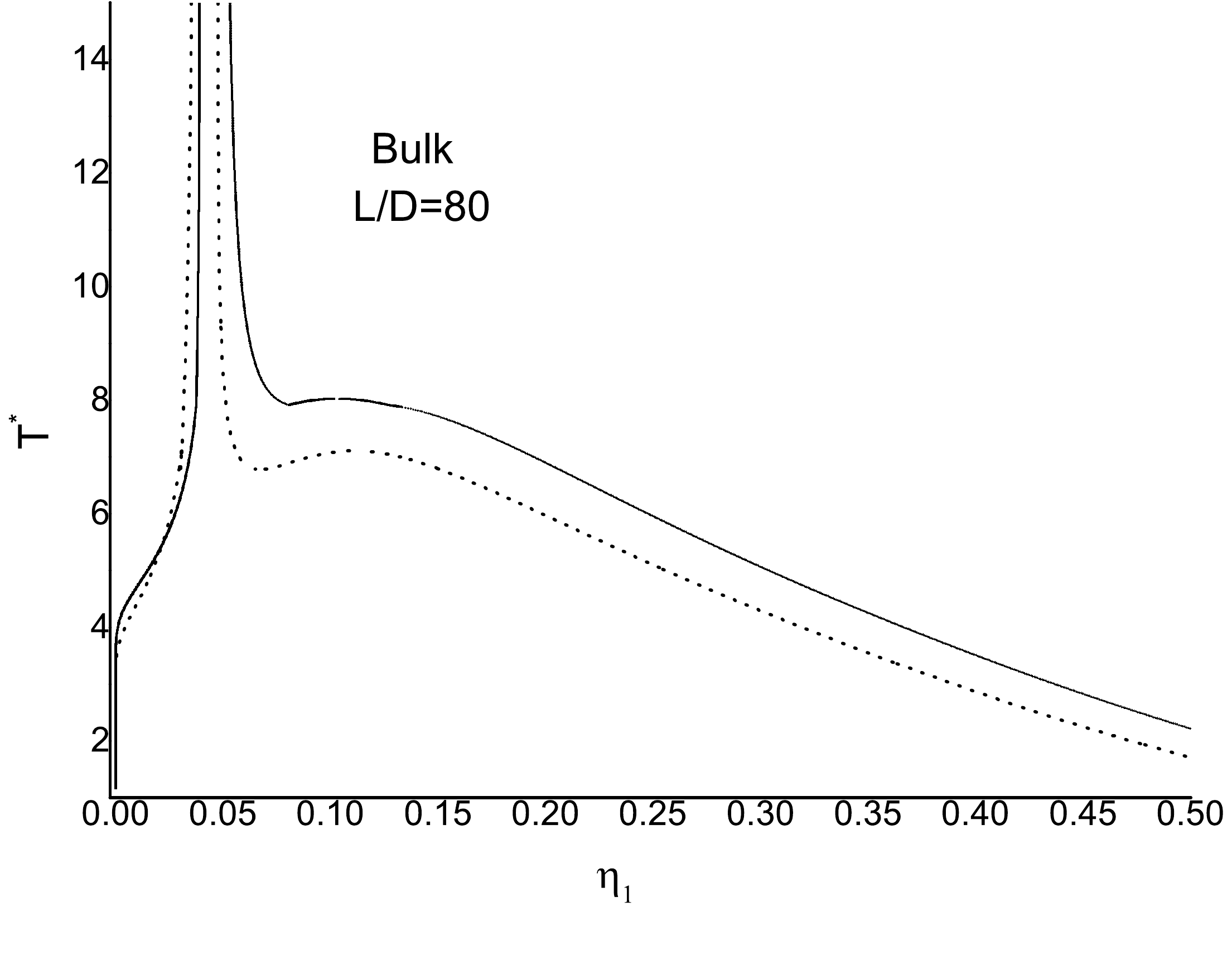}
	}
	\caption{Comparison of the phase diagram for a hard-spherocylinder fluid with the modified anisotropic Lennard-Jones attraction (solid line) and that with an anisotropic Lennard-Jones attraction in the isotropic approximation (\ref{hol_smot3.4}) for the attractive contribution (dotted line).}
	\label{Fig2}
\end{figure}
\begin{figure}[!t]
	\centerline{
				\includegraphics [height=0.45\textwidth]{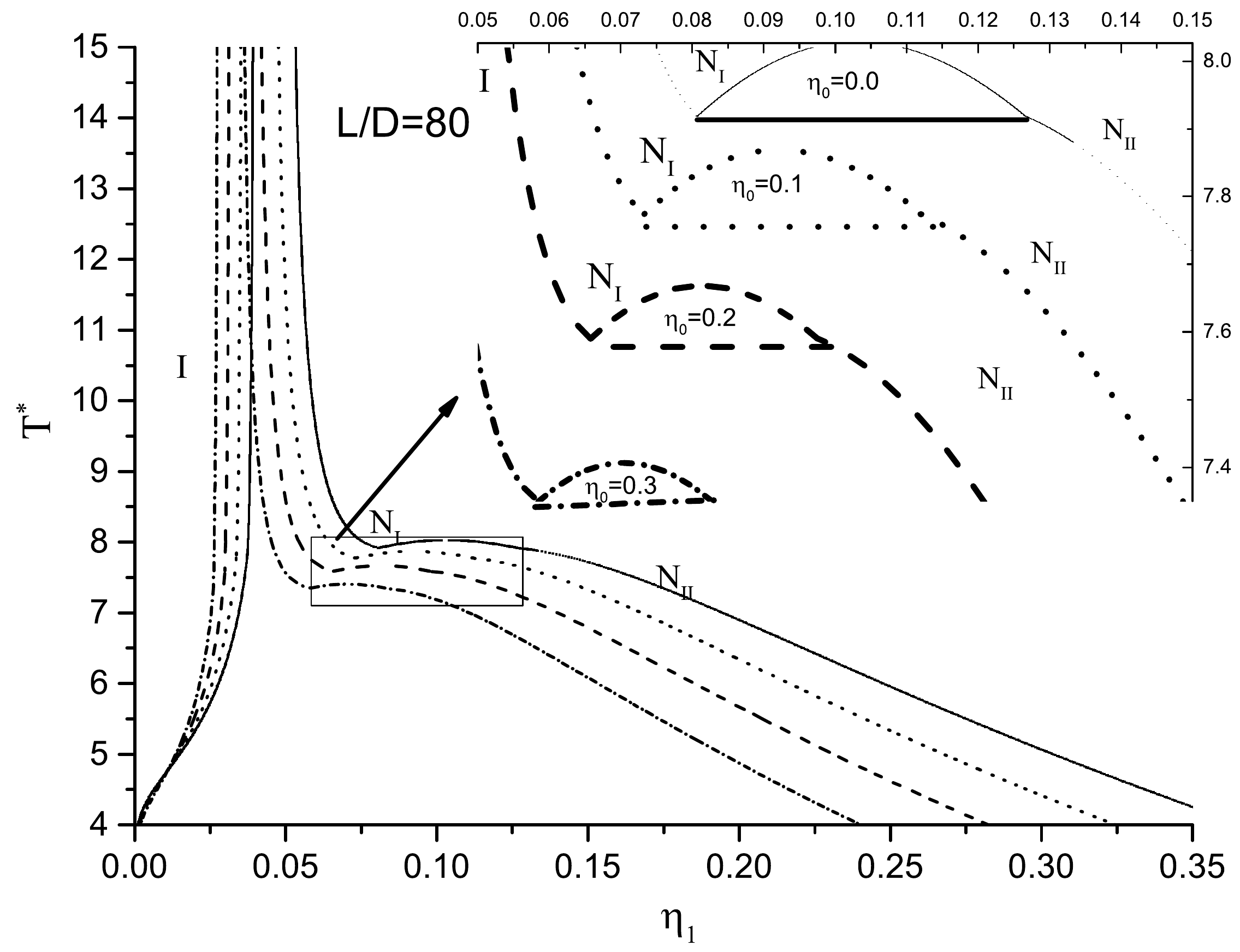}
	}
	\caption{Phase diagrams for a hard spherocylinder fluid with the modified anisotropic Lennard-Jones attraction in a disordered porous medium.}
	\label{Fig3}
\end{figure}

The phase diagram for a hard spherocylinder fluid with a Lennard-Jones-like attraction modified in the form (\ref{hol_smot3.5}) for the bulk case for ${L_{1}}/{D_{1}}=80$ is presented in figure \ref{Fig2}. This is a typical phase diagram with an isotropic-nematic transition with coexistence of two nematic phases with different densities. For comparison, in figure \ref{Fig2} the phase diagram for the same model of a hard spherocylinder fluid with the Lennard-Jones attraction in the form (\ref{hol_smot3.2}) with simplification (\ref{hol_smot3.4}) for the excluded volume is also presented. We cannot see a big difference between phase behaviors in these two cases.

The influence of porous media on the phase behavior of a hard spherocylinder fluid with Lennard-Jones-like attraction in the form (\ref{hol_smot3.5}) in the bulk case and the presence of porous media modelled by randomly distributed hard spheres with the packing fractions $\eta_{0}=0.1,0.2,0.3$ is illustrated in figure \ref{Fig3} for the case ${L_{1}}/{D_{1}}=80$. We can expect the appearance of two nematic states with different densities. A decrease of matrix porosity $\phi_{0}=1-\eta_{0}$ (or an increase of packing fraction $\eta_{0}$) shifts this transition to lower densities and to lower temperatures.
 
However, the model with a modified Lennard-Jones attraction similar to the model with the original Lennard-Jones attraction has the same anisotropy of attraction for different distances only due to $r^{*}={r}/{\sigma\left(\Omega_{1},\Omega_{2},\Omega_{r}\right)}$. In a real situation, parallel configuration of two spherocylinders has the largest attraction. Thus, a simple model was formulated to incorporate such an attraction \cite{Vroege92,GrosKho81} --- an attractive square-well potential around the hard core of spherocylinders, the so-called anisotropic square-well potential, in which the potential $u^\text{attr}\left(r,\Omega_{1},\Omega_{2}\right)$ has the form
\begin{align}
u^\text{attr}\left(r,\Omega_{1},\Omega_{2}\right)&=\left\{\begin{array}{ll}
-\epsilon_{0}\,,&
\gamma_{1}D_{1}>r>\sigma\left(\Omega_{1},\Omega_{2},\Omega_{r}\right),\\
0\,,& r< \sigma(\Omega_{1},\Omega_{2},\Omega_{r}),\,\, r>\gamma_{1}D_{1}\,.
\end{array}\right.
\label{hol_smot3.7}
\end{align}

This is the third model which we use in this paper for the description of phase behavior of a hard spherocylinder fluid with anisotropic attraction. We note that this potential was used in G. Jackson's group for the investigation of the phase behavior of a hard spherocylinder fluid using the Onsager trial function for the singlet orientational distribution function $f\left(\Omega_{1}\right)$ \cite{Wu,Franco-Melgar}.

After application of the potential (\ref{hol_smot3.7}) for the attractive constant $a$ we have
\begin{equation}
a=-\frac{\epsilon_{0}}{\phi_{0}V_{1}}\left[\frac{4}{3}\piup\gamma_{1}^{3}D_{1}^{3}-\int \rd\Omega_{1}\rd\Omega_{2}f\left(\Omega_{1}\right)f\left(\Omega_{2}\right)V_{1}^\text{exc}\left(\Omega_{1},\Omega_{2}\right)\right].
\label{hol_smot3.8}
\end{equation}
\begin{figure}[!b]
	\centerline{
				\includegraphics [height=0.45\textwidth]{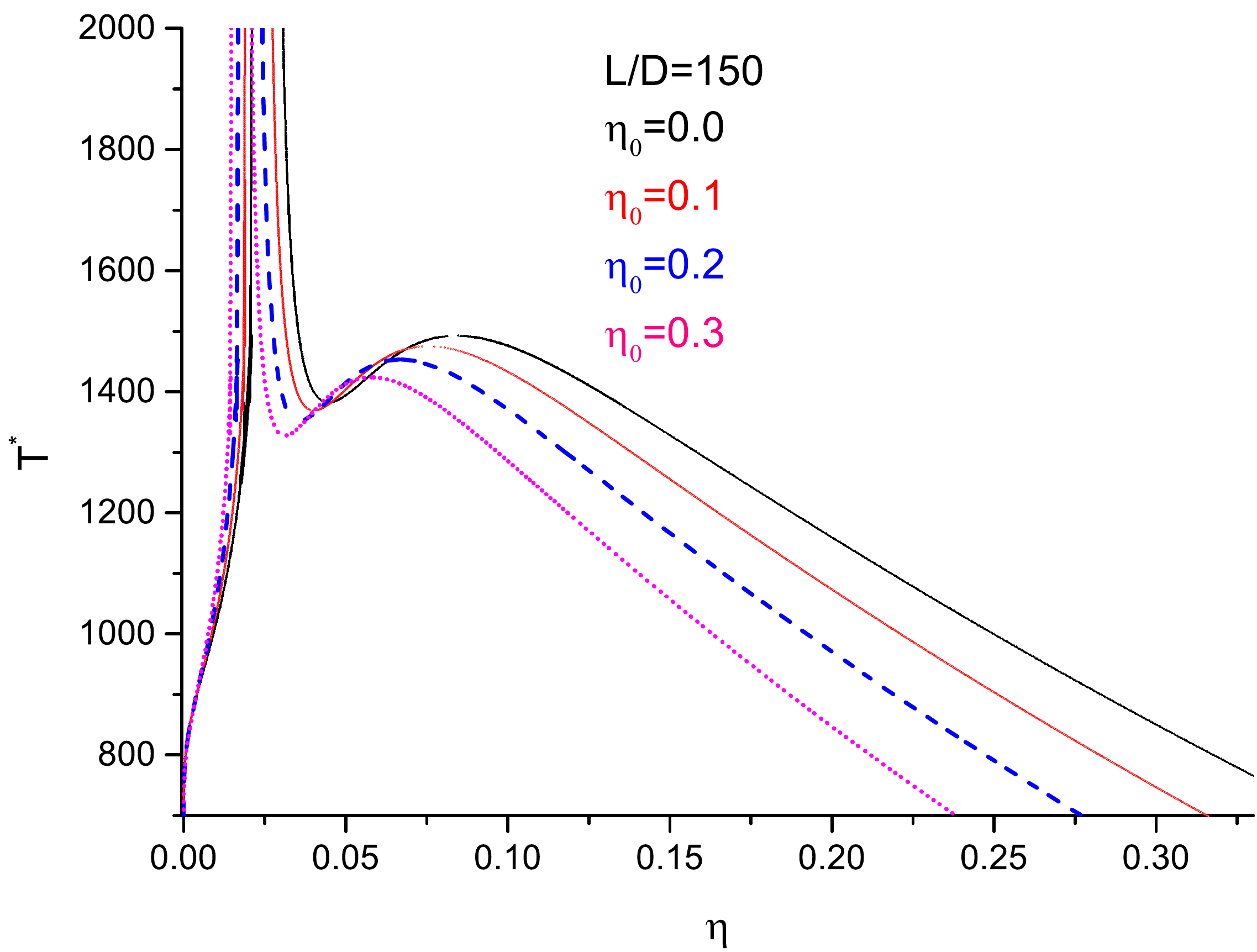}
	}
	\caption{(Colour online) Phase diagrams for a hard spherocylinder fluid with an anisotropic square-well attraction in a disordered porous medium.}
	\label{Fig4}
\end{figure}

The phase diagram for a hard spherocylinder fluid with attraction in the form (\ref{hol_smot3.7}) in the bulk case and in the presence of porous media at ${L_{1}}/{D_{1}}=150$ is presented in figure \ref{Fig4}. As we can see, the phase diagrams for the case considered and for a hard spherocylinder fluid with a modified Lennard-Jones attraction are very similar qualitatively but the temperatures of transition between two nematic phases are two or three orders higher in the case of anisotropic square-well attractive potential compared with the modified Lennard-Jones potential. In both models, a decrease of porosity shifts the nematic-nematic transition to lower densities and to lower temperatures.

\begin{figure}[!b]
	\centerline{
				\includegraphics [height=0.45\textwidth]{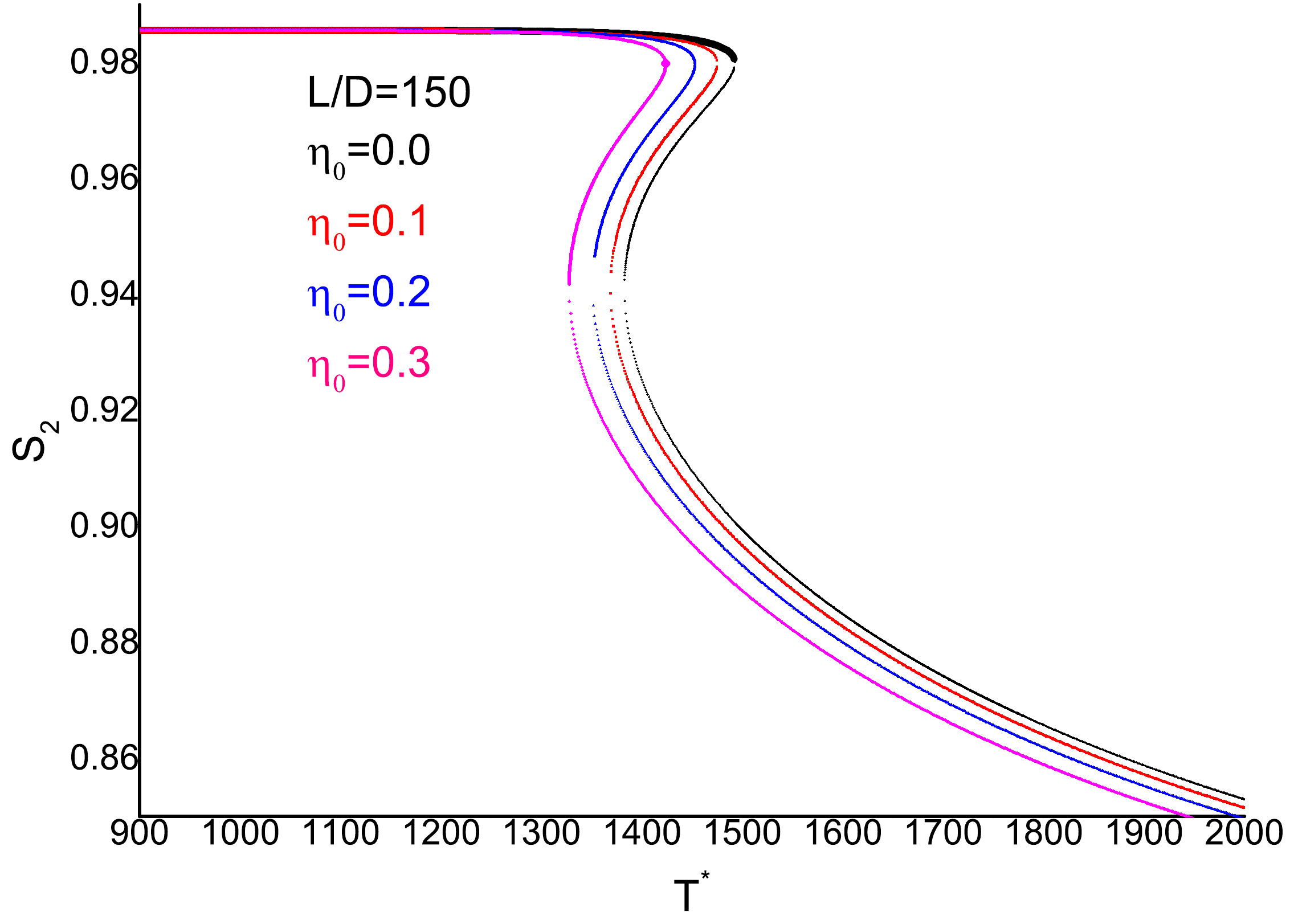}
	}
	\caption{(Colour online) The temperature dependence of the nematic order parameter in coexisting nematic phases for a hard spherocylinder fluid with an anisotropic square-well attraction in a disordered porous medium.}
	\label{Fig5}
\end{figure}
Finally, in figure \ref{Fig5} the temperature dependence of the nematic order parameter
\begin{equation}
S_{2}=\int f\left(\Omega\right)P_{2}\left(\cos\vartheta\right)\rd\Omega
\label{hol_smot3.9}
\end{equation}
in coexisting nematic phases for a hard spherocylinder fluid with anisotropic square-well attraction is presented. As we can see, the orientational ordering in a high density nematic phase $N_{2}$ is near $1$ in the entire temperature region. In the low-density nematic phase $N_{1}$ the order parameter $S_{2}$ monotonously increases with the temperature. In the critical point $T_\text{c}$ both curves meet each other. With an increasing porosity, the order parameter decreases.

\section{Conclusions}

In this paper we present the improved version of the generalized Van der Waals equation for anisotropic fluids in disordered porous media formulated by us in our previous paper \cite{HolShmotPat,HolShmot2015}. As usual, in the Van der Waals approach the expressions for thermodynamic properties of a fluid have two different parts. The first one is connected with hard core repulsive interactions and is based on analytical expressions for the equation of state and for the chemical potential of a hard spherocylinder fluid in random porous media obtained in the framework of the scaled particle theory. In particular, in the present paper the analytical expressions for the pressure and the chemical potential of a hard spherocylinder fluid in a random porous medium obtained in the SPT2b1 approximation with Carnahan-Starling and Parsons-Lee corrections were used \cite{HolShmot2018}. The second term of the generalized Van der Waals equation is connected with the mean value of the attractive interparticle interaction. The important improvement of the generalized Van der Waals equation is connected with taking into account the coupling between repulsive and attractive contributions in the treatment of attractive interparticle interaction. We note that this important aspect was neglected in our previous papers \cite{HolShmotPat,HolShmot2015}. From minimization of the free energy of the fluid we obtained a nonlinear integral equation for the singlet distribution function which describes the orientational ordering in the fluid. Due to the coupling between attractive and repulsive contributions, the excluded volume $V_{1}^\text{exc}\left(\Omega_{1},\Omega_{2}\right)$ introduces the terms corresponding to the isotropic and anisotropic contributions. In the simple case when in the potential of interparticle interaction $U^\text{attr}\left(r_{12},\Omega_{1},\Omega_{2}\right)$ the term corresponding to Maier-Saupe interaction $\left(\epsilon_{2}=0\right)$ is absent, the integral equation for the singlet distribution function has the same form (\ref{hol_smot2.47}) as a respective equation for a hard spherocylinder fluid obtained by Onsager \cite{Onsager49}. However, a corresponding constant $C$ in this equation is presented as the sum of two terms connected with repulsive and attractive contributions.

In this paper we focused on the consideration of spherocylinders with rather large elongations, for which the isotropic-nematic phase transition appears in the region of quite small densities. Due to this, the corresponding gas-liquid transition is located completely in the nematic region and can be treated as a nematic-nematic transition. We showed that the phase diagram is very sensitive to the type of attractive interparticle interaction. We consider three such simple models with $\epsilon_{2}=0$, namely hard spherocylinders with Lennard-Jones-like attraction in the form (\ref{hol_smot3.2}), hard spherocylinders with the modified Lennard-Jones attraction in the form (\ref{hol_smot3.5}) and hard spherocylinders with an anisotropic square-well attraction in the form (\ref{hol_smot3.7}). The phase diagram for the first model with Lennard-Jones attraction presented in figure \ref{Fig1} is considerably different from the results \cite{HolShmotPat,HolShmot2015} obtained in the framework of the isotropic approximation~(\ref{hol_smot3.4}) for the excluded volume $V_{1}^\text{exc}\left(\Omega_{1},\Omega_{2}\right)$ formed by two spherocylinders. More or less similar results with \cite{HolShmotPat,HolShmot2015} for the phase behaviour were obtained for the modified version of Lennard-Jones attraction with cutting and shifting of Lennard-Jones attraction at distances ${r}/{\sigma\left(\Omega_{1},\Omega_{2},\Omega_{r}\right)}=\gamma_{1}=1+{L_{1}}/{D_{1}}$. It means that the original Lennard-Jones model significantly overestimates the role of attraction at distances larger than $\gamma_{1}$. The third model is the model with the anisotropic square-well potential (\ref{hol_smot3.7}). This model manifests a qualitatively similar phase behaviour as the model with the modified Lennard-Jones attraction. Both models demonstrate the existence of a nematic-nematic phase separation but the temperatures of this transition are two or three orders higher for the model with the square-well attraction than in the case of the modified Lennard-Jones version. The temperature of transition significantly increases with increasing lengths of spherocylinders. A similar effect was not observed for the model with modified Lennard-Jones attraction. It was shown that the higher-density nematic phase is more ordered than the lower-density nematic phase. In both models a decrease of porosity (or an increase of packing fraction of matrix particles) shifts the nematic-nematic transition to lower densities and to lower temperatures. With an increasing porosity the ordering in both nematic phases decreases.

Finally, we note that all the expressions in this paper up to equation (\ref{hol_smot2.40}) are presented really in the high temperature approximation (HTA), which includes the pair distribution function of the reference system $g_{2}^\text{hs}\left({r}/{\sigma}\right)$. In this paper we present this pair distribution function in the low-density limit~(\ref{hol_smot2.42}) which corresponds to Van der Waals approach and all calculations following this were done in this approximation. We note that in our previous paper \cite{HolPatSmot2015} to describe the phase behavior of a simple fluid we used the HTA approach, and the pair distribution functions for the hard sphere fluid in a random porous medium $g_{2}^\text{hs}\left({r}/{\sigma}\right)$ were obtained from the numerical solution of the replica Ornstein-Zernike (ROZ) equation for this model. The results obtained demonstrate good aggrement between the calculated liquid-vapour phase diagrams of a Lennard-Jones fluid in a hard sphere matrix and the corresponding computer simulation data. To describe the pair distribution function of the reference system, the interpolation scheme \cite{HvozdKalyu2017} can also be used which combines the contact value obtained from the SPT theory with the analytical results for the pair distribution function in the bulk case with an effective density. In our future studies we plan to extend such approaches to the case of anisotropic fluids in random porous media. In addition, we plan to use the theory developed for interpretation of liquid-crystalline states of polypeptide solutions and other biological systems in porous media \cite{Vaf2010}.

\section{Acknowledgements}
This project has received funding from Ministry of Education and Science of Ukraine (project No~M/116-2019) and V.S. acknowledges the support of the NAS of Ukraine (young scientists research works project No 30-04/09-2019).

We thank Ivan Kravtsiv for the careful reading of the manuscript and useful comments.

%
%

\ukrainianpart

\title{Узагальнення підходу Ван дер Ваальса для ізотропно-нематичної фазової рівноваги в анізотропних плинах у невпорядкованому пористому середовищі}
\author{М.Ф. Головко, В.І. Шмотолоха}
\address{Інститут фізики конденсованих систем НAH України, вул. Свєнціцького, 1, 79011 Львів, Україна
}
%
%
%

\makeukrtitle

\begin{abstract}
\tolerance=3000%
Пропонується узагальнення підходу Ван дер Ваальса для анізотропних плинів у невпорядкованих пористих середовищах.
В якості системи відліку розглядається твердий сфероциліндричний плин у невпорядкованому пористому середовищі, що
описується в рамках теорії масштабної частинки з поправками Карнахана-Старлінга і Парсонса-Лі.
Притягальна частина взаємодії розглядається в рамках наближення середнього поля, в якому враховується орієнтаційна
залежність виключеного об'єму двох сфероциліндрів. Ми зосереджуємося на розгляді сфероциліндричного плину з досить довгою довжиною,
в якому встановлено нематично-нематичний фазовий перехід. Показано, що ці дві нематичні фази мають різні
густини і характеризуються різним орієнтаційним впорядкуванням. Встановлено значний вплив форми міжчастинкової притягальної взаємодії
на фазову поведінку анізотропних плинів у невпорядкованих пористих середовищах. Цей ефект продемонстровано на прикладі трьох простих моделей, а саме моделі з анізотропним притяганням Леннарда-Джонса, моделі з модифікованим притяганням
Леннарда-Джонса і моделі з анізотропним потенціалом квадратної ями. Для всіх розглянутих моделей
зменшення пористості зсуває фазову діаграму до області нижчих густин і нижчих температур.

\keywords {тверді сфероциліндричні плини, невпорядковане пористе середовище, теорія масштабної частинки, узагальнення рівняння Ван дер Ваальса, ізотропно-нематична рівновага, нематично-нематичний перехід}
\end{abstract}
\end{document}